\documentclass[10pt]{iopart}
\usepackage{psfig}
\newcommand{\be}{\begin{equation}}
\newcommand{\ee}{\end{equation}}

\newcommand{\bea}{\begin{eqnarray}}
\newcommand{\eea}{\end{eqnarray}}

\newcommand{\lton}{\mathrel{\lower.9ex
                  \hbox{$\stackrel{\displaystyle <}{\sim}$}}}

\begin{document}

\title[News on Strangeness at Ultrarelativistic Energies - Review 
of Microscopic Models]
{News on Strangeness at Ultrarelativistic Energies - Review of 
Microscopic Models}

\author{Sven Soff$\,^{1,2}$\footnote[3]{ssoff@th.physik.uni-frankfurt.de}} 

\address{$^1$ Nuclear Science Division 70-319, 
Lawrence Berkeley National Laboratory, 
One Cyclotron Road, Berkeley, CA94720, USA}

\address{$^2$ Institut f\"ur Theoretische Physik, Goethe-Universit\"at, 
Postfach 111932, 60054 Frankfurt am Main, Germany}

\begin{abstract}
We review recent developments in the field of microscopic transport model 
calculations for ultrarelativistic heavy ion collisions.
In particular, we focus on the strangeness production, for example, 
the $\phi$-meson and its role as a messenger of the early phase of
the system evolution. Moreover, we discuss the important effects 
of the (soft) field properties on the multiparticle system.
We outline some current problems of the models as well as possible 
solutions to them.  
\end{abstract}
\vspace{-0.5cm}
\section{Introduction}
The purpose of this overview is to present recent developments 
in the field of microscopic transport theory for 
ultrarelativistic heavy ion collisions, that is, 
for energies at SPS and RHIC. 
Special emphasis is put on the strangeness production.
Moreover, we will particularly focus on the role of the $\phi$-meson 
as well as on the importance of the color field properties 
on {\it strange} observables.
Furthermore, we will also briefly address the topics baryon transport, 
excitation functions, multi-particle collisions, initial conditions, 
$J/\Psi$ data, and particle interferometry.
 
\section{Color Fields}
An example of what is meant by the importance of field effects on the 
dynamics is given by the 
quark molecular dynamics (qMD) 
model \cite{Scherer:2001ap,Hofmann:1999jx}. 
It is used to study microscopically
the dynamics of colored quarks. 
The intention is to get an effective description of the non-perturbative, 
soft gluonic part of QCD.
In particular, the hadronization process itself can be described explicitly.
The initial quark distributions are obtained, for example, 
from the UrQMD model.
The non-equilibrium dynamics of hadronization and the loss
of correlations among quarks can be studied subsequently by qMD.
The semi-classical model
has a two-body color potential between the (anti)quarks. 
In addition, one has to define a dynamical hadronization
criterion. 
The quarks 
interact via a Cornell potential with color matrices. 
Thus, they carry color, (flavor, spin, and isospin).
The simple model Hamiltonian is
\[
H =
\sum_{i=1}^N\sqrt{p_i^2+m_i^2}+\frac{1}{2}\sum_{i,j}C_{ij}
V(\left\vert\mathbf{r}_i-\mathbf{r}_j\right\vert)
\]
where $N$ are the number of quarks in the system. 
The potential $V(r)$ is linearly increasing at large distances (confinement)
and has a Coulomb-type behavior at short distances. 
The Cornell-potential is    
\[
V(r) = -\frac{3}{4}\frac{\alpha_s}{r}+\kappa\,r\;.
\]
A further approximation is done by allowing only the {\it 
diagonal colorless} gluon exchange, that is, the quarks always keep their 
color. 
The color matrix elements $C_{ij}$ determine the sign and
relative strength of the interaction depending on the color 
combination of the pair 
(for details see \cite{Scherer:2001ap}).

The rather simple model is surprisingly able to describe a 
large amount of experimental data, for example, 
various particle spectra and ratios. 
Concerning strangeness, we show here the interesting feature of 
aquiring net strangeness in the quark phase. 
According to the idea of the strangeness distillation 
antistrange quarks hadronize earlier
in a baryon rich system. Fig.~1 shows the number of 
net strange quarks (s-$\bar{{\rm s}}$) as a function of time 
for the baryon rich system in central Pb+Pb collisions at 
30$\,A$GeV. Values up to approximately 10 are reached depending on the 
string tension. This indicates the possibility to study 
hyperon matter, hypernuclei, or even strangelets in these high 
baryon density collisions around the future GSI energies.
\parbox{7.8cm}{
\psfig{figure=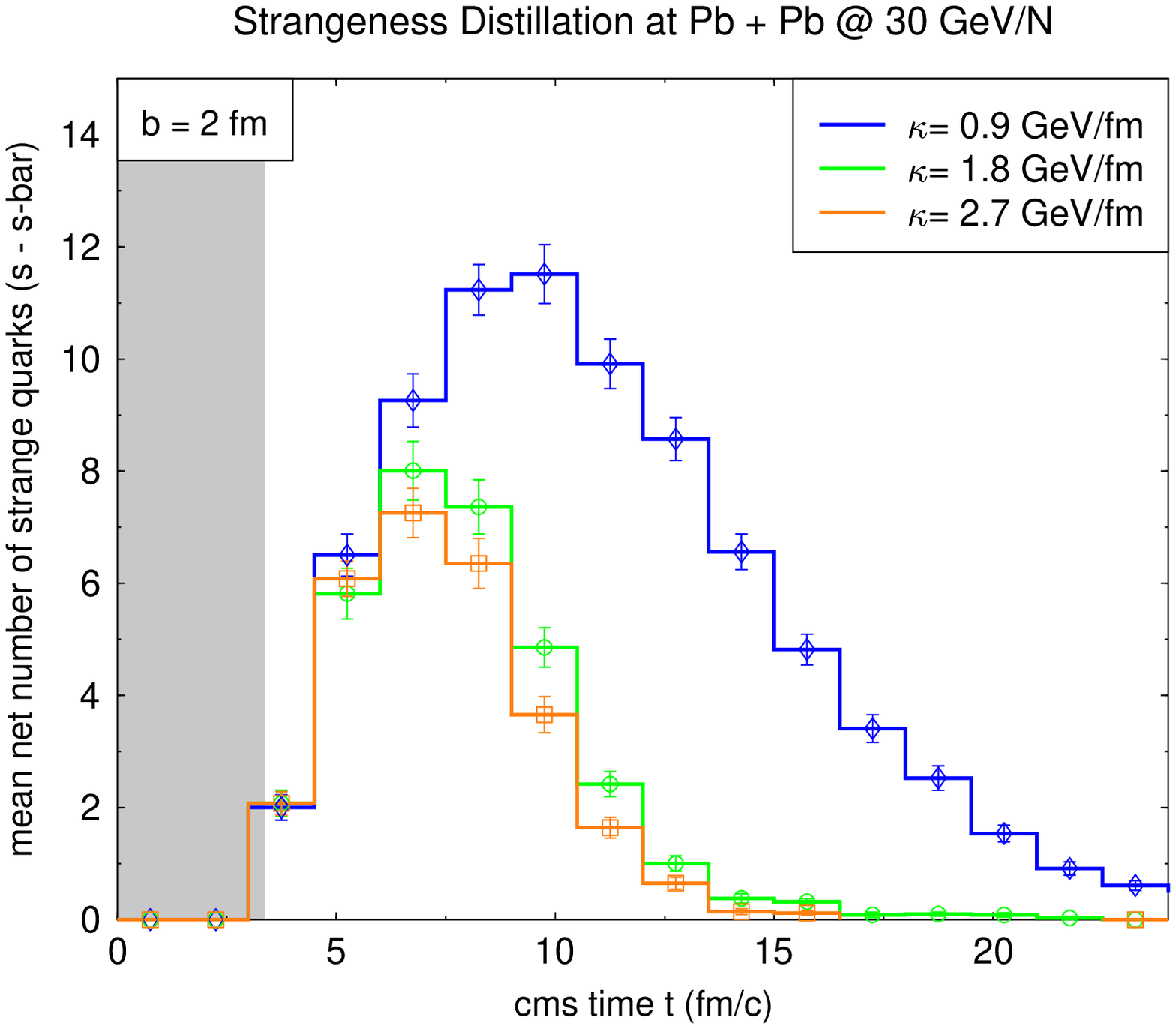,width=7.cm}
}
\parbox{5cm}{{\small Figure 1:
qMD calculation for Pb+Pb at 30$\,A$GeV by S.~Scherer
\protect{\cite{Scherer:2001ap}}.
The number of net strange quarks is shown as a function of time for
different values of the string tension $\kappa$.
}
}

Another feature of the model is to study the details of the 
parton-hadron phase transition that is mainly driven by quark rearrangement.
Late stage quark-antiquark pair production is a rather rare process 
in this model because the strong color fields that are needed are 
screened.
Fig.~2 shows the distribution of the mean path length of quarks 
before hadronizing. Quarks from the same original hadron 
(solid line) recluster faster than quarks from different original 
hadrons (dotted line).
The hadron formation follows an exponential decay of the quark 
cluster to hadrons. 
The reclustering exhibits a different {\it decay length} than the 
rearrangement process (2.2 vs.\ 4.8\,fm). 
The amount of reclustering ($\approx 50\%$ for S+Au) 
demonstrates the survival of correlations in the 
quark system and possibly incomplete thermalization. 
This kind of detailed analysis also provides additional insight 
to the question of the mechanisms of hadronization (recombination vs.\ 
fragmentation \cite{Fries:2003rf}).
 
\parbox{7.3cm}{\psfig{figure=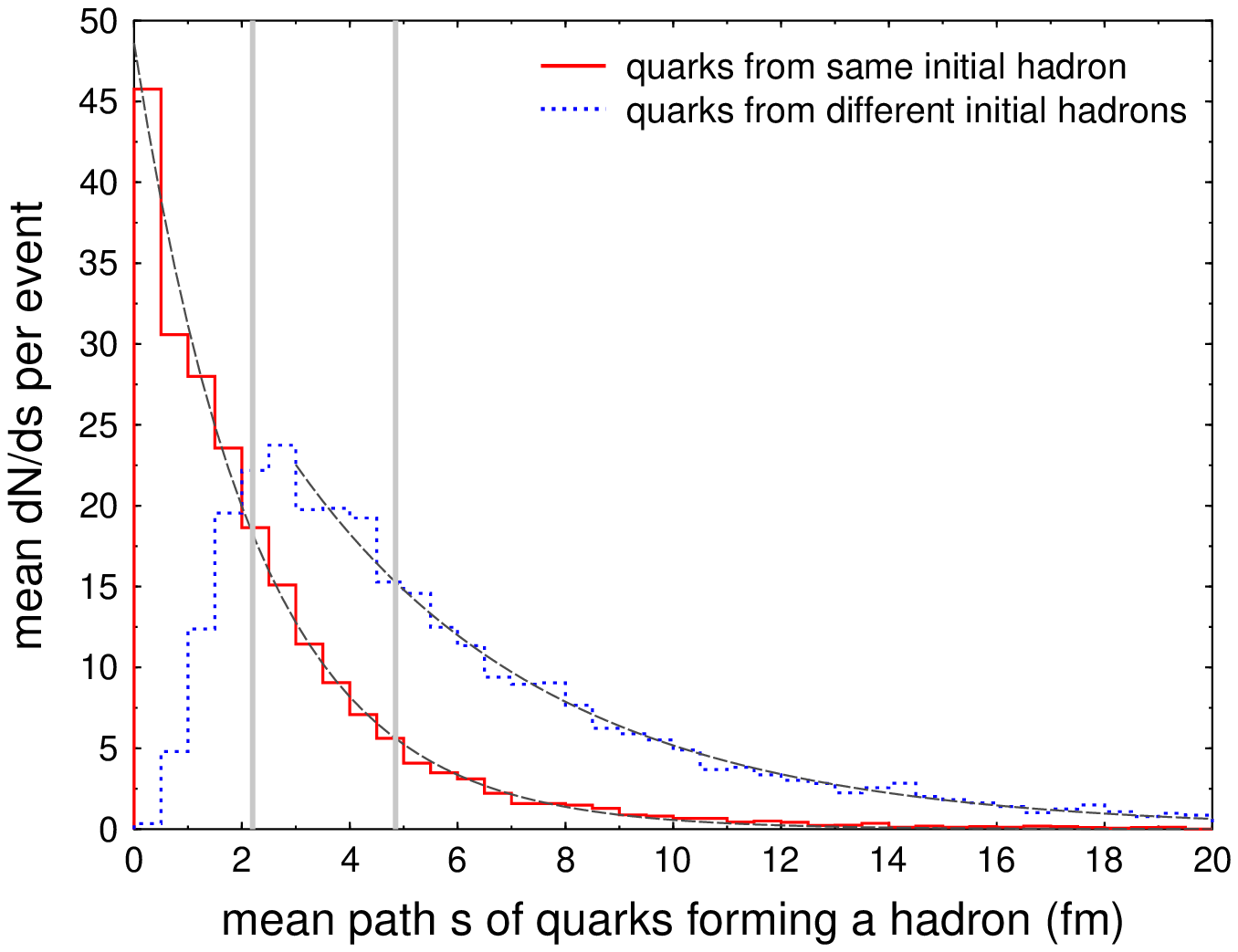,width=7.0cm}}
\parbox{5cm}{{\small Figure 2: Quark Molecular Dynamics simulation of 
the hadronization in S(200\,$A$GeV)Au collisions. Quarks originating from 
the same hadron hadronize after a shorter path length $s$ than 
quarks steming from different hadrons. Figure taken 
from \protect{\cite{Scherer:2001ap}}.
}
}


\section{$\phi$-mesons, baryon transport 
and the sensitivity to field properties}

The field properties also enter in the description of heavy ion 
collisions by means of cascade calculations that are based on 
hadron, (di)quark, and string degrees of freedom, as for example 
the UrQMD model \cite{urqmd}. 
Here the initial particle production is dominated by 
string excitations and fragmentations, that is, longitudinally 
excited color flux tubes. 
In a dense colored medium the color field strength might 
be considerably enhanced. 
The particle formation process in hadronic collisions
can be viewed as quantum tunneling of quark-antiquark and gluon pairs
in the presence of a background color electric field.
It is formed between two receding hadrons which are color charged
by the exchange of soft gluons while colliding.
In nucleus-nucleus collisions the color charges may be considerably greater
than in nucleon-nucleon collisions due to the almost simultaneous
interaction of several participating nucleons.
With increasing energy of the target and projectile the number and density of
strings grows, so that they start overlapping, forming clusters,
which act as new effective sources for
particle production \cite{biro84,sor92}.
It has been predicted that the multiplicities
of, for example, strange baryons or antibaryons should be
strongly enhanced \cite{sor92,Soff:1999et} once the color
field strength grows.
The abundances of (multiply) strange (anti)baryons
in central Pb+Pb collisions at Cern-SPS \cite{and98a}, for example,
can only be explained within
the framework of microscopic model calculations \cite{Soff:1999et}
if the elementary production probability of $s\overline{s}$ pairs, which
is governed in the string models \cite{Andersson:ia}
by the Schwinger mechanism \cite{schwing51}
$\sim \exp (- \pi m_q^2/2\kappa)$,
is considerably enhanced.
This corresponds either to
a dramatic enhancement of the string tension $\kappa$
(from the default $\sim 1\,$GeV/fm to $3\,$GeV/fm) or to
quark masses $m_q$ that are reduced from their constituent quark values
to current quark values as motivated by chiral symmetry
restoration.
In a fully analog way, $\overline{p}$ abundances can be explained
since an enhanced string tension similarly leads to an
increased production probability of antidiquark-diquark pairs which
is needed to account for the experimentally observed
yields \cite{Bleicher:2000gj}.
However, this argument, providing additional motivation,
has to be reconsidered if multifusion processes \cite{rapp},
e.g., $5\pi \rightarrow \overline{B}B$, that are neglected
in microscopic transport models based on $2\rightarrow n $ scatterings,
contribute significantly to the baryon pair production.
A variation of the string tension from $\kappa=1\,$GeV/fm to $3\,$GeV/fm
increases the pair production probability of
strange quarks (compared to light quarks)
from $\gamma_s=P(s\overline{s})/P(q\overline{q})=0.37$ to 0.72.
Similarly, the diquark production probability is enhanced
from $\gamma_{qq}=P(qq\overline{q}\overline{q})/P(q\overline{q})=0.093$
to 0.45. In general, heavier flavors or diquarks ($Q$) are suppressed
according to the Schwinger formula \cite{schwing51} by
\[
\gamma_Q=\frac{P(Q\overline{Q})}{P(q\overline{q})}= \exp\left(
-\frac{\pi (m_Q^2-m_q^2)}{2 \kappa} \right)\,.
\]

The effectively enhanced string tension in
a densely colored environment also follows from its relation
to the Regge slope $\alpha'$. In the rotational string picture
the string tension $\kappa$ is related to the Regge slope
$\alpha'$ by \cite{goddard,Wong:jf}
\[
\kappa=\frac{1}{2 \pi  \alpha'}.
\label{alphakappa}
\]
The empirical value of the Regge slope for baryons is
$\alpha'\approx 1\,$GeV$^{-2}$ \cite{green87} that yields
a string tension of approximately $1\,$GeV/fm.
However, high-energetic processes dominated by Pomeron exchange,
characterizing
the multi-gluon exchange processes existent in high-energetic nucleus nucleus
collisions, are described by a Regge trajectory (Pomeron)
with a smaller slope of $\alpha_P'\approx 0.4\,$GeV$^{-2}$
\cite{veneziano,collins}.
According to Eq.\ (\ref{alphakappa}) this translates into a
considerably larger ({\it effective}) string tension $\kappa$.
  
Strong color fields also have an impact on the formation times of 
newly produced (di)quarks and thus on collision rates. 
The formation times are inversely proportional to the
string tension 
\[
t_f \sim 1/ \kappa.
\]
The larger the string tension the shorter and short-living are the
strings with a certain total energy. 

\vspace*{-0.4cm}
\parbox{12cm}{
\psfig{figure=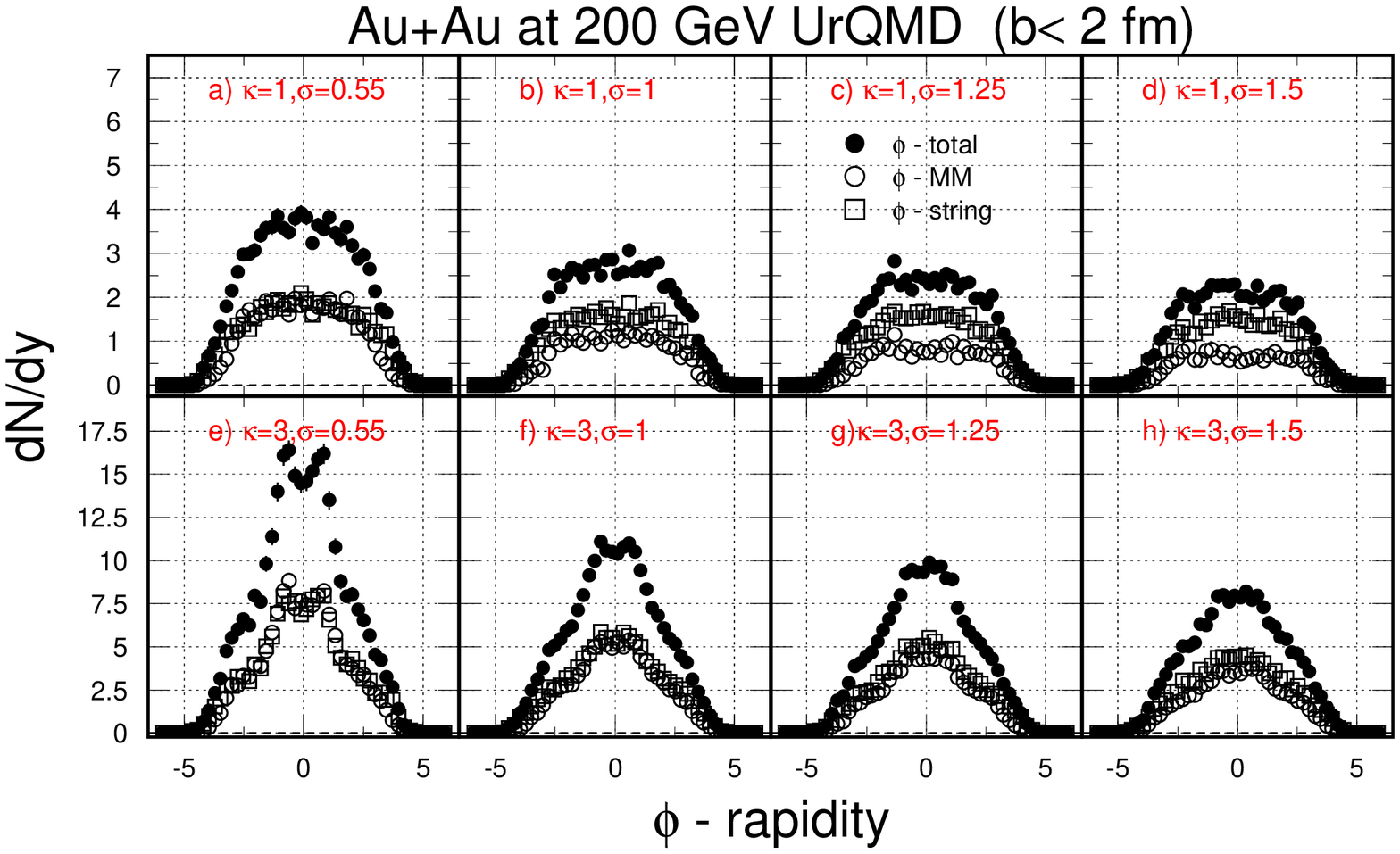,width=10.cm}
}

\vspace*{-0.55cm}
\parbox{12cm}{{{\small Figure 3: Rapidity distributions of
$\phi$-mesons in central Au+Au collisions at RHIC.
Besides the distributions of all $\phi$'s (full circles),
the contributions from $\phi$'s produced in (resonant) meson-meson
(K$\bar{{\rm K}}$) collisions (open circles) and from $\phi$'s originating
from string decays (open squares) are shown.
Each panel corresponds to different values of the
string tension ($\kappa$) (mass term) and the intrinsic transverse momentum
broadening parameter $\sigma$.}}
}

Moreover, Schwinger's formula does not only apply to the mass 
term but to the transverse momentum term, too. 
While $\kappa$ determines the production probability, 
the intrinsic transverse momentum parameter 
$\sigma$ defines the transverse momentum scale. 
The vacuum value of $0.55\,$GeV/c (width of a Gaussian) 
may be considerably enhanced 
by the strong color fields. 
Here, we vary both parameters independently to study systematically their 
individual effects on the observables \cite{soffneu}. 
Fig.~3 shows the rapidity spectra of $\phi$-mesons 
at RHIC. The intrinsic $p_t$ is varied from 0.55 to 1.5 GeV/c (left to right). 
The upper and lower panels correspond to the vacuum string tension 
and the SCF scenario, respectively.
Clearly a larger $\kappa$ value leads to a strongly enhanced 
$\phi$ production. 
Increasing the intrinsic $p_t$ decreases the 
$\phi$-meson yield. 
The reason for this becomes immediatley clear when looking 
at the production channels. The production in meson-meson 
(K$\bar{\rm K}$) collisions gets considerably reduced when 
$\sigma$ is enhanced. 
The coalescence-like production loses due to a growing phase space. 
The experimental study of the $\phi$-meson in both decay channels 
and the observed differences (due to rescattering effects) 
at RHIC will help  to disentangle the different source contributions 
and thus enlight the role of strong color fields.  

Another observable that has been shown to be highly sensitive to 
SCFs is the netbaryon number transport. Prior to the collision it 
is non-vanishing only at beam and target rapidities ($\Delta y \sim 11$ units 
for RHIC!).
Recent data by the STAR \cite{Adler:2001aq},
PHOBOS \cite{Back:2001qr},
BRAHMS \cite{Bearden:2001kt}, and
PHENIX \cite{Adcox:2001mf} collaborations,
show a considerable number of netprotons ($p-\overline{p}$) at midrapidity. 
and antibaryon-to-baryon ratios smaller than unity.

The calculated anti-baryon-to-baryon ratios at midrapidity are shown in Fig.~4
as a function of the strangeness content $|S|$, both
for $\kappa=1\,$GeV/fm as well as for the strong color
field scenario (SCF, $\kappa=3\,$GeV/fm) \cite{Soff:2002bn}.
The $\overline{B}/B$-ratios increase with the strangeness content $|S|$ of
the baryons and also increase with with impact parameter $b$.
There is stronger absorption of antibaryons in central collisions (reducing
the numerator of the ratio) and more {\it stopping}
(increasing the denominator).
Most important, the $\overline{B}/B$-ratios also increase
with the color field strength $\kappa$ what is basically due to an increased 
pair production (while more collisions should reduce the ratios).
The SCF results are closer to data. Also the observed  
{\it hyperonization} in the netbaryon distributions 
supports the SCF scenario (the 
$(\Lambda-\overline{\Lambda})/(p-\overline{p})$ ratio is $\sim 0.8$
for the strong color field (SCF) ($\kappa=3\,$GeV/fm) scenario, corresponding
to the experimental value \cite{Adcox:2001mf,Adler:2001aq} but smaller 
for the {\it vacuum} calculations. 

\hspace*{-1cm}
\parbox{7.4cm}{
\psfig{figure=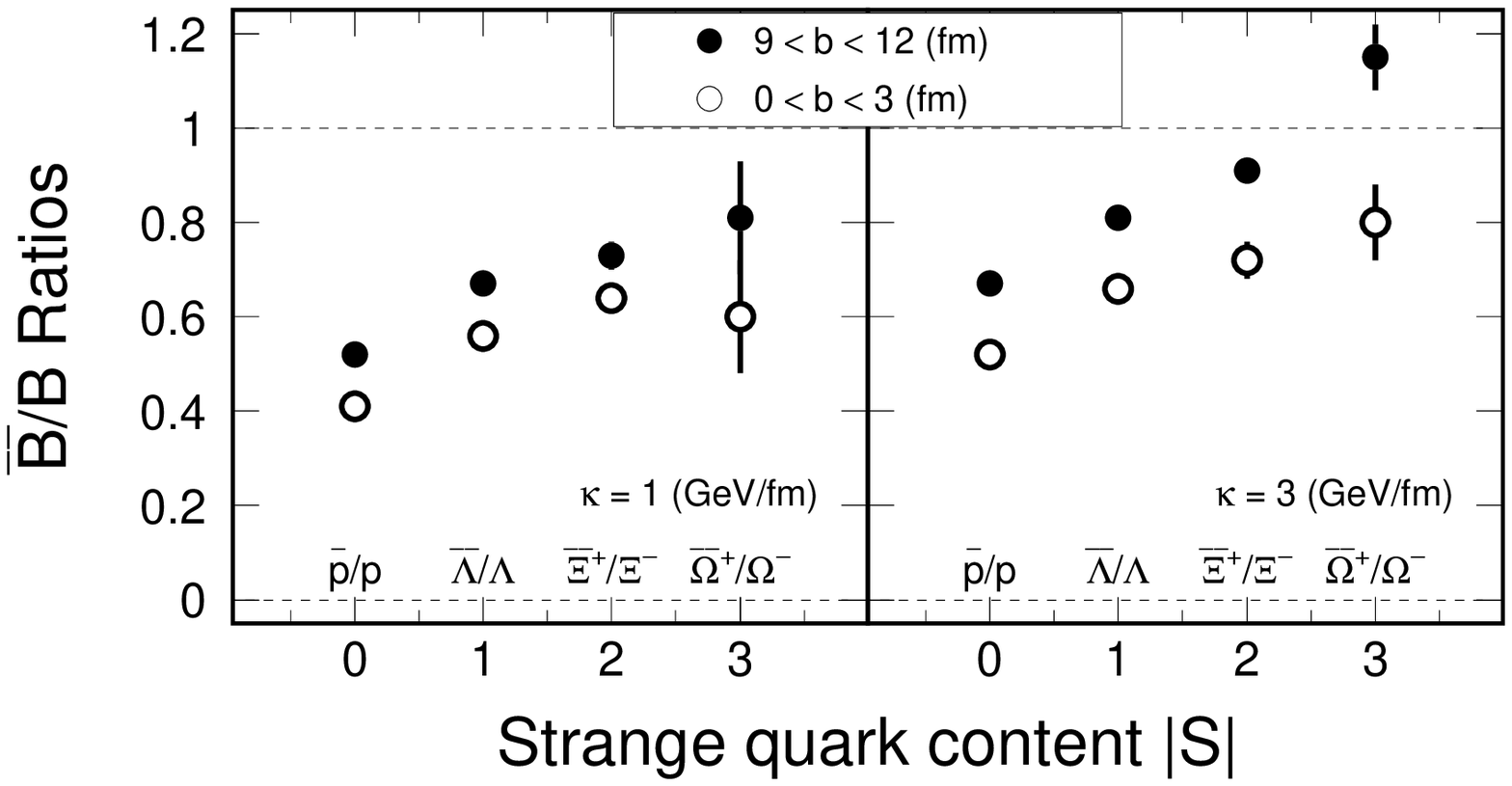,width=8cm}
}
\parbox{5.2cm}{{\small Figure 4: 
Antibaryon-to-baryon ratios at midrapidity as a function
of the strangeness content $|S|$ in
Au+Au collisions at RHIC ($\sqrt{s}_{NN}=200\,$GeV). Calculations with
a string tension of $\kappa=1\,$GeV/fm are shown on the left
and the results with strong color fields ($\kappa=3\,$GeV/fm)
are shown on the right \protect{\cite{Soff:2002bn}}.} 
}

A complementary or alternative {\it ansatz} to describe the 
baryon transport is provided through the parton cascade model (PCM).
Fig.~5 shows the netbaryon distributions as calculated with the  
VNI/BMS version \cite{Bass:2002vm,Bass:2002fh} of the parton cascade. 
It is interesting to note that a considerable fraction of 
the observed netbaryon number at midrapidity is already contained 
in the parton distribution functions of the original nuclei. 
Other contributions arise from secondary scatterings of the partons and 
fragmentation that together yield a netbaryon number compatible with 
data. 
It will be interesting to see whether the correct 
longitudinal dynamics will also imply the strong transverse expansion 
dynamics as, for example, the relatively large elliptic flow. 

\parbox{7.8cm}{
\psfig{figure=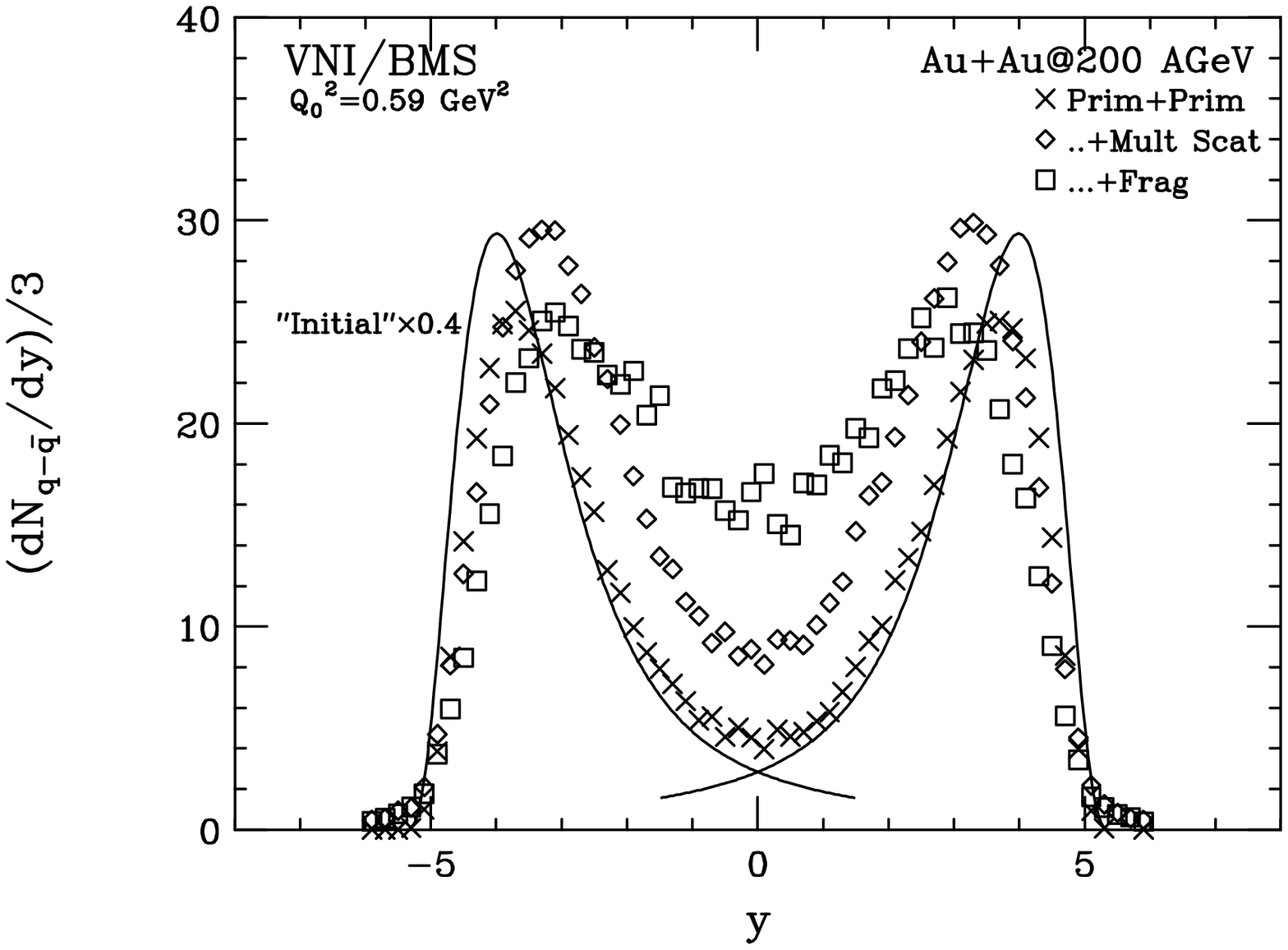,width=7cm}
}
\parbox{5cm}{{\small Figure 5: 
Netbaryon rapidity distributions for Au+Au at RHIC.
Crosses are calculation with only primary parton
scatterings, rhombes include parton rescattering and squares include
rescattering and parton fragmentation. The solid lines show the
netbaryon content of the partonic distribution functions for gold
nuclei, scaled by an average liberation factor of 0.4. 
Figure taken from \cite{Bass:2002vm}. 
}
}

The production of $\phi$ mesons in Au+Au collisions at RHIC 
has also been studied with the quark gluon string model 
(QGSM) \cite{Bravina:2002qg}.
It is found that the inverse slope parameter agree with data but  
the absolute yield of $\phi$'s is underestimated
by a factor 2. Somehow surprising, the fusion of strings does not 
increase the $\phi$ yield in QGSM.
Similar to previous UrQMD results \cite{Soff:1999et} the early production is
governed by string
processes. The resonant kaon rescattering production channel 
dominates in the later stages. 
Fig.~6 illustrates
the freeze-out distribution of the $\phi$ mesons over rapidity and
emission time from QGSM.  

\vspace*{-1.6cm}
\parbox{7.8cm}{
\psfig{figure=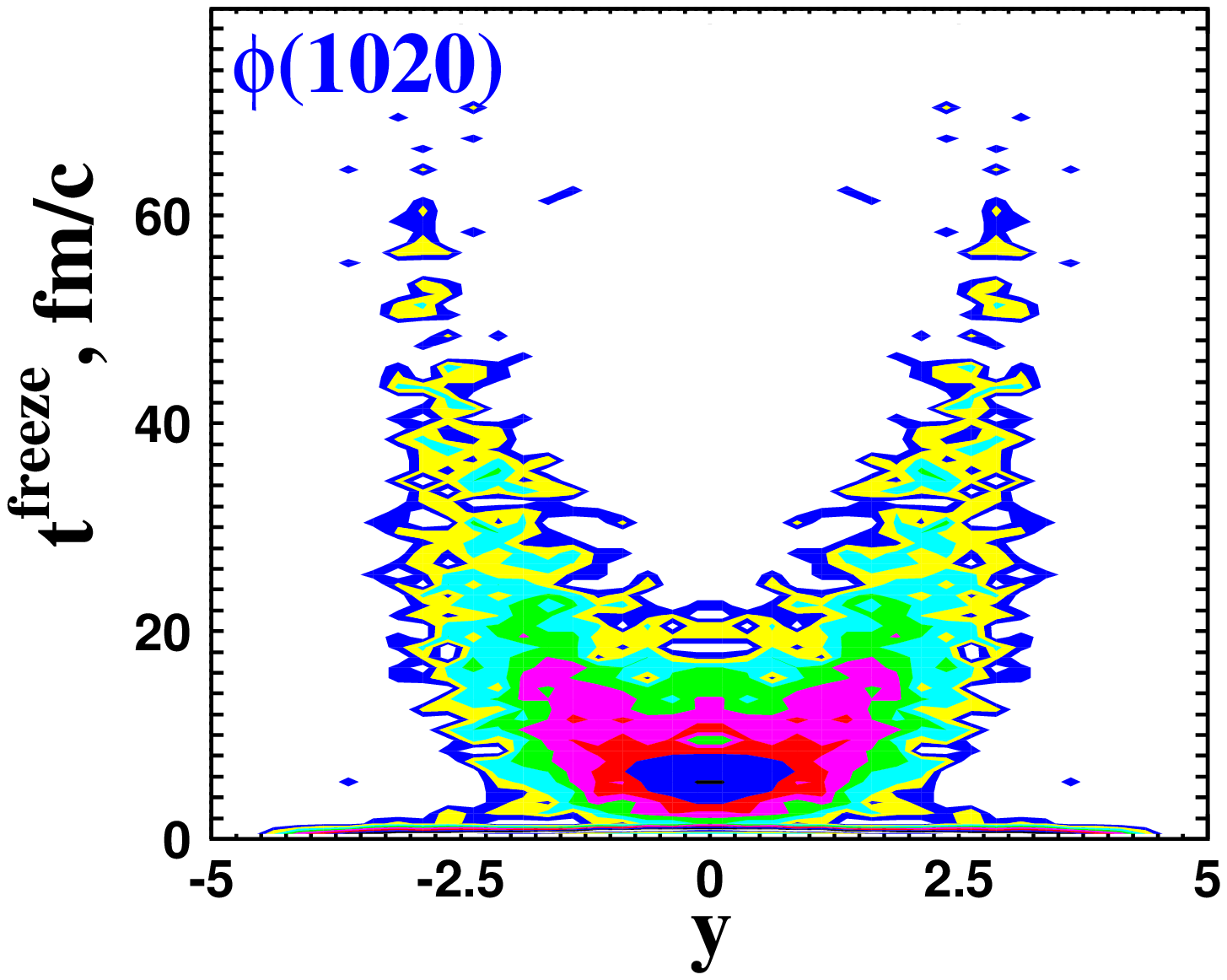,width=7.5cm}
}
\parbox{5cm}{{\small Figure 6: 
$d^2N/dy dt$ distribution of the final-state $\phi$ mesons in the
$(y,t)$-plane from QGSM \protect{\cite{Bravina:2002qg}}
}
}
\vspace*{-1cm}

The $\phi$ puzzle has also been addressed with a systematic study 
of the kaon and dimuon decay channels by the AMPT model \cite{Pal:2002aw}. 
The AMPT model is a multiphase transport model 
that includes the initial partonic and final
hadronic interactions. It starts with minijet 
partons and
strings from the HIJING mode.
The partons enter the ZPC parton cascade model. The parton hadron transition 
is based on the Lund string
fragmentation model. Final-state hadronic
scatterings are modeled by the ART model.
Similar to the other models the resonant K$\bar{\rm K}$ production channel 
also contributes considerably to the total $\phi$ yield. 
However, even with in-medium modifications of the kaons and $\phi$'s 
the observed differences in the yields cannot be fully explained 
(see figure 7). 
An additional ad hoc production is needed to come close to the 
data (of NA50). It is important to remark that the initial 
$\phi$'s (those that are possibly created through strong color 
fields) will be primarily visible in the dimuon channel.

\hspace*{-0.7cm}
\parbox{7.8cm}{
\psfig{figure=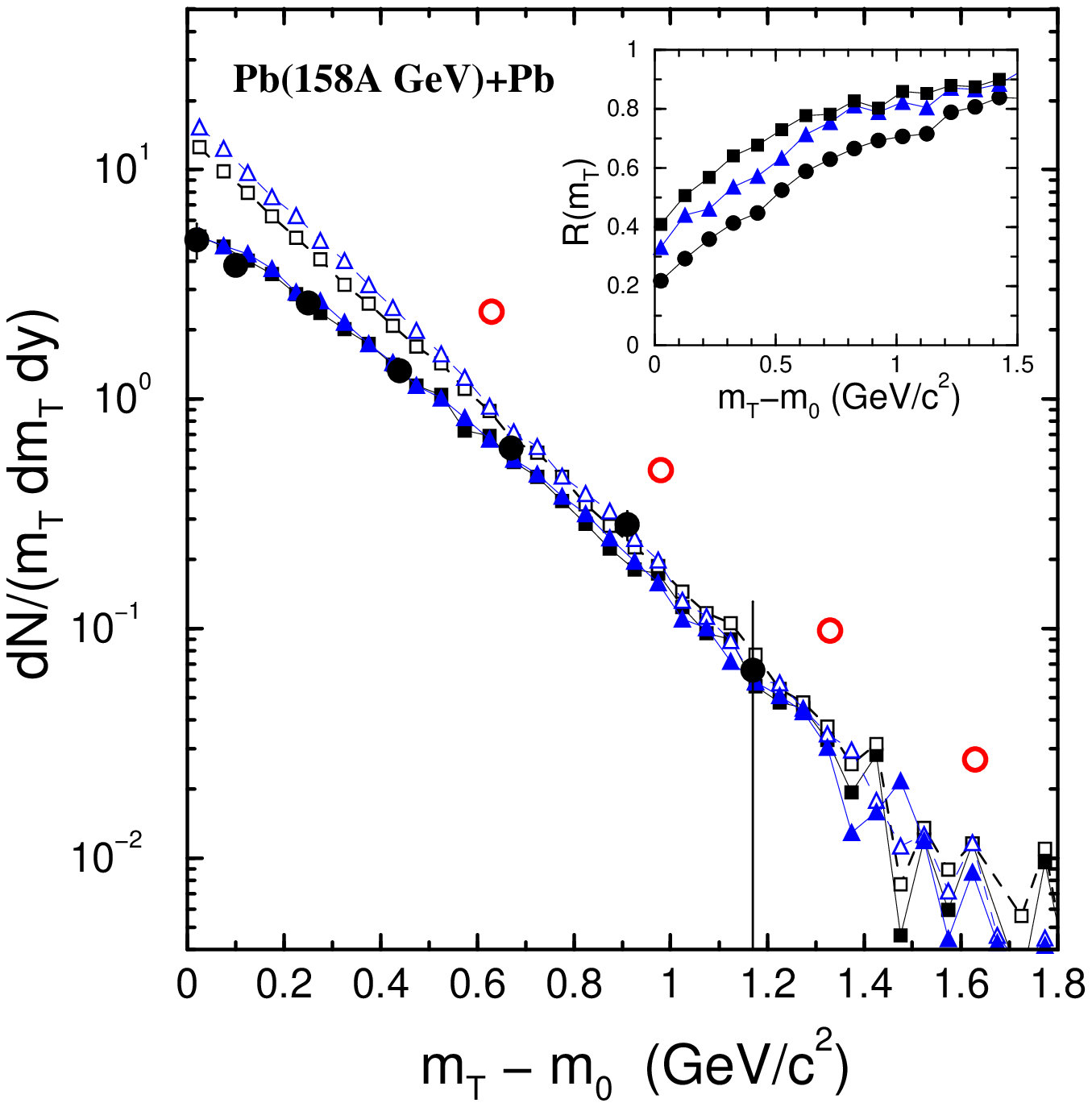,width=6.8cm}
}
\parbox{5.2cm}{{\small Figure 7: 
Transverse mass spectra of $\phi$'s 
reconstructed from $K^+K^-$ pairs (solid symbols) and from $\mu^+\mu^-$
channel (open symbols) for Pb+Pb at $158A$ GeV 
($b\leq 3.5$, $|y|<1$) fm in the AMPT model. The results are for
without (squares) and with (triangles) in-medium mass modifications.
The solid circles are NA49 data \protect\cite{na49} for
$\phi\to K^+K^-$ decay, and the open circles are NA50 data
\protect\cite{friese} for $\phi\to \mu^+\mu^-$.
In the inset is shown as a function of $m_T$ the ratio $R(m_T)$ for
phi mesons decaying to kaon-antikaon pairs that are not scattered
to those determined from the dimuon channel. 
Circles are for an artificially increased number of 
$\phi$'s by a factor 2 in 
the HIJING model. Figure from \protect{\cite{Pal:2002aw}}.
}
}

\section{Further Topics}
\subsection{Excitation functions of meson abundancies and multi-meson 
fusion}

New data from the Cern/SPS allow us to study the excitation function 
of particle production (and spectra, see \cite{Gazdzicki:2003fj}) 
in the interesting energy range of maximum netbaryon densities. 
Characteristic features in the data as maxima in the 
K$^+$/$\pi^+$ ratio gave reason to speculate about the observation 
of the phase transition. 
How do 'conventional' hadron transport theories compare 
to data \cite{prc,WBS_K02,Reiter:2003zj}?
The detailed comparison with two realisations of the cascade model 
with hadron resonance, (di)quark, and string degrees of freedom 
(UrQMD and HSD) shows that the kaon yields are described reasonably 
(see Fig.~8). 
However, the pion yields and hence the K$^+$/$\pi^+$ ratios are off 
(see figure 9). 

\vspace*{0.4cm}
\parbox{12cm}{
\psfig{figure=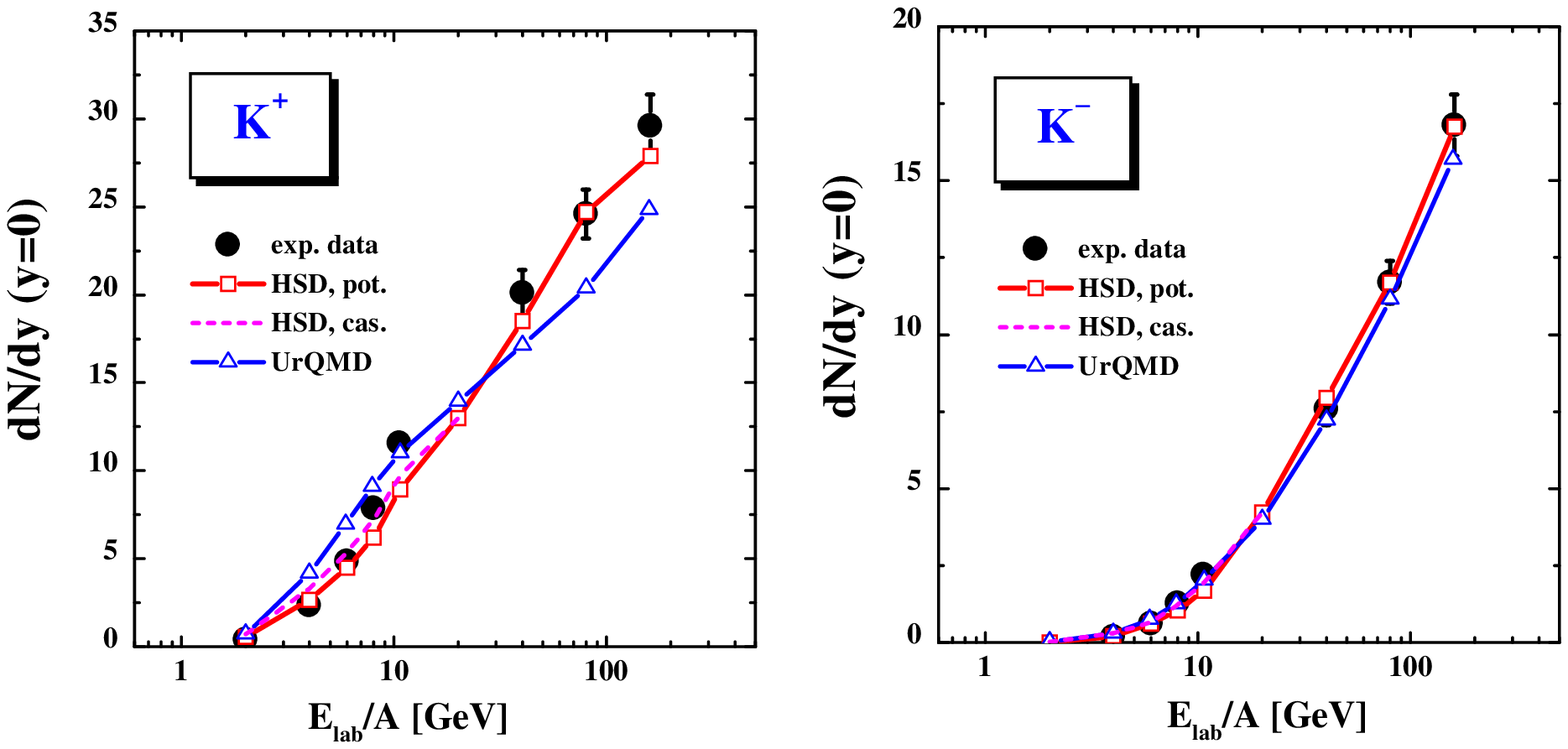,width=10cm}
}

\parbox{12cm}{{\small Figure 8:
Excitation function of $K^+$ and $K^-$ yields from central Au+Au (AGS)
or Pb+Pb (SPS) collisions  \protect\cite{prc} in comparison to experimental
data~\protect\cite{NA49_new,E866E917}.
}
}
\vspace*{0.15cm}

The overestimation of the pion yields in the model calculations
could be due to a lack of multiparticle back reactions in the
model that also lead to 'shorter' chemical equilibration times. 
$2\rightarrow n$ collisions (with $n>2$) are performed but
the inverse channels are not taken care of.
Another possible explanation is given through enhanced
pion in-medium masses.
In fact, the importance of
$3\leftrightarrow 2$ transitions has been explored in an extended HSD
transport approach~\cite{Cassing:2001ds} for antiproton production by
meson fusion in $A + A$ collisions at the AGS and SPS 
(see figure 10). In order to achieve a
more conclusive answer from transport studies multi-particle interactions
deserve further investigation in future generations of transport codes.

\vspace*{0.2cm}
\parbox{12cm}{
\psfig{figure=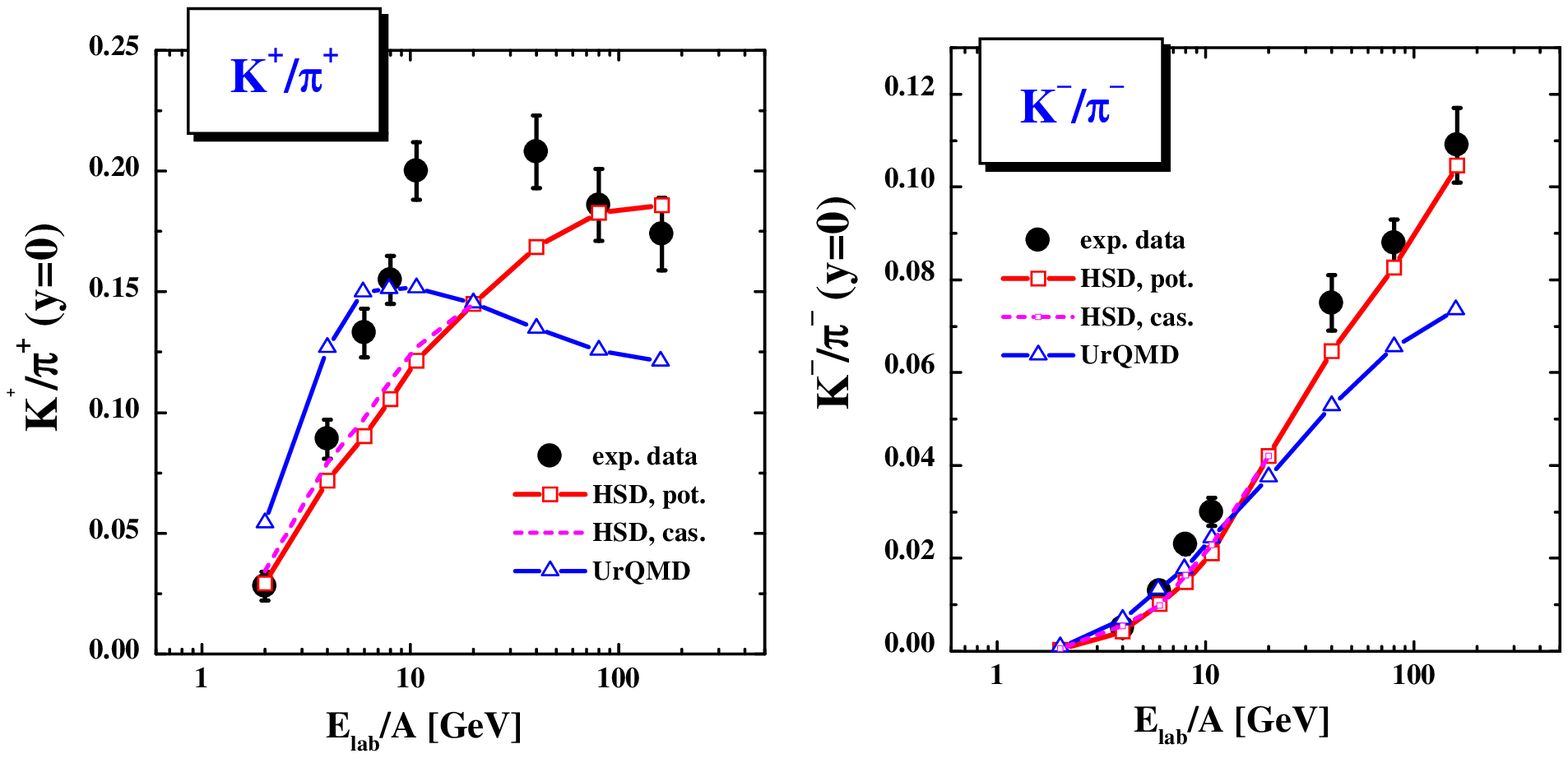,width=10cm}
}

\parbox{12cm}{{\small Figure 9: 
The same as in Fig.~8 for the $K^+/\pi^+$ and $K^-/\pi^-$ ratios. 
}
}

\parbox{12cm}{
\psfig{figure=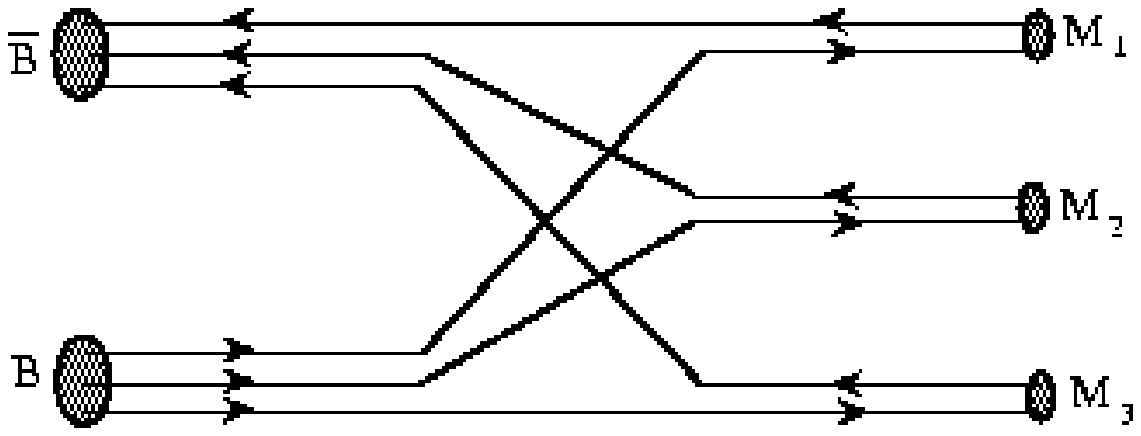,width=10.cm}
}

\parbox{12cm}{{\small Figure 10: 
Illustration of the flavor rearrangement model for
$B\bar{B}$ annihilation to 3 mesons and vice versa. The mesons
$M_i$ may be either pseudo-scalar or vector mesons, respectively. 
Figure from \protect{\cite{Cassing:2001ds}}
}
}

Within this approach 
the abundancies
of  antiprotons as observed from peripheral to central collisions
of $Pb + Pb$ at the SPS and $Au + Au$ at the AGS can approximately
be described. 

Fig.~11 shows the reaction rate $B+\bar{B}
\rightarrow mesons$ vs.\ the backward reaction rate 
for central Pb+Pb collisions at 160 $A$GeV. 
Both rates turn out to be 
comparable within the statistics demonstrating an approximate local
chemical equilibrium. The final $\bar{p}$'s origin dominantly 
($\sim$ 88\%) from the 3 meson fusion reactions. 
These studies indicate that these back reactions may have 
dramatic effects and therefore ask for further investigations. 

\parbox{7.8cm}{
\psfig{figure=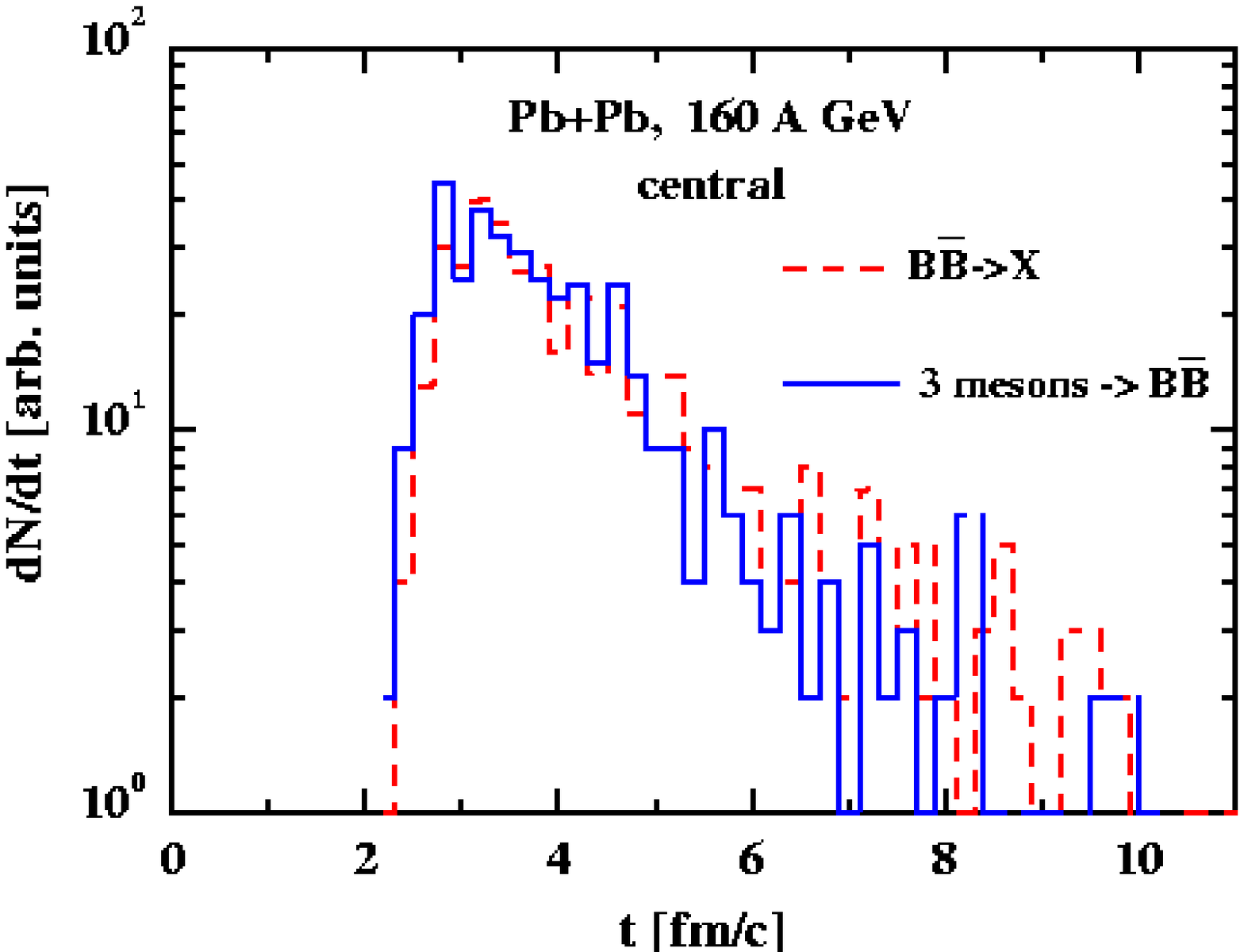,width=7cm}
}
\parbox{5cm}{{\small Figure 11: 
The annihilation rate $B\bar{B} \rightarrow $ mesons (dashed histogram)
for central $Pb+Pb$ collisions at 160 $A\,$GeV as a function
of time in comparison to the backward reaction rate (solid
histogram) within the HSD transport approach. Figure 
from \protect{\cite{Cassing:2001ds}}.}}

\subsection{J/$\psi$}

Another possible QGP signal is charmonium production 
and its anomalous suppression. 
It is worth to mention that the new data 
from NA50 \cite{ramello}, however, do not deviate 
significantly from the predictions of standard-type 
cascade calculations (UrQMD or HSD) 
\cite{Spieles:1999kp,Cassing:1997kw} (Fig.~12). 
Both the
$E_{\rm T}$ dependence and the $E_{\rm ZDC}$ dependence do no longer 
support the observation of a strong anomalous suppression. 

\parbox{12cm}{
\psfig{figure=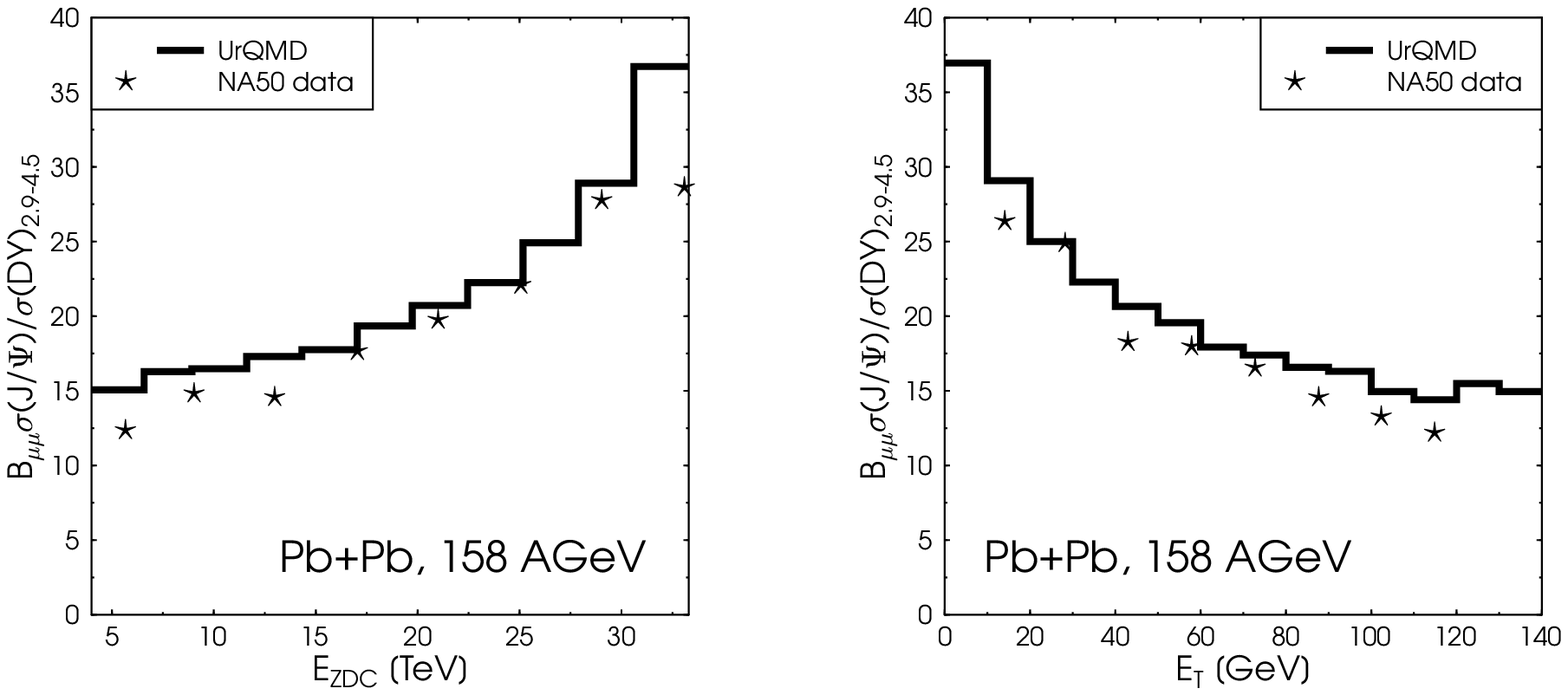,width=10.cm}
}

\parbox{12cm}{{\small Figure 12: 
The ratio of $J/\Psi$ to Drell-Yan production as a function of forward
energy $E_{\rm ZDC}$ (left) and transverse energy $E_{\rm T}$ (right) for 
Pb+Pb at 158 AGeV. The preliminary data from the
year 2000 NA50 Pb+Pb run~\protect\cite{ramello} are compared to  
UrQMD results~\protect\cite{Spieles:1999kp}. Figure from \protect\cite{Reiter:2003zj}}
}

\subsection{Initial conditions}
The newly developed NEXUS3 model addresses the initial conditions 
of pp or AA collisions \cite{Drescher:2000ha,kwerner}. 
It provides a self-consistent description of the energy sharing 
among the participating partons in the collision. 
The idea is called parton based Gribov-Regge theory, that is, 
soft and hard processes are described by (multi-)pomeron exchange. 
As one consequence strings are not necessarily attached to 
the valence quarks but may be connected to sea quarks.  
An observable supporting this picture may be  given by the 
$\bar{\Omega}/\Omega$ ratio in pp collisions which is smaller or 
close to unity for NEXUS3 but larger for standard string models 
(or the older versions of NEXUS) \cite{Bleicher:2001nz}.

\subsection{Particle interferometry}
In brief, progress has been made at several frontiers. 
More realistic models were developed and used to calculate 
the multidimensional correlation functions. 
Hybrid-type models, for example, that allow one to study 
explicitly the sensitivities to the QCD equation of state by 
simultaneously describing the freeze-out in a realistic way 
were used to discuss the so-called RHIC HBT-puzzle. 
The impact of in-medium modifications on HBT results were discussed 
for the first time. 
For a discussion and references, see for 
example \cite{Soff:2002pc,kaonlett,soffbassdumi}.  

\section{Outlook}
Topics not discussed here are, for example, the necessary tests 
of the microscopic models by simulating infinite matter. 
The upcoming RHIC pA data will put severe constraints for the models. 
An exact understanding of the elliptic flow $v_2$ is needed 
(collision numbers, formation times).   
How is the (partially) hydrodynamic behavior realized on a microscopic level? 
Also, there has been quite some progress in the charm sector, 
for example, the prediction of a substantial production of $J/\Psi$'s 
in D$\bar{\rm D}$ collisions \cite{Bratkovskaya:2003ux}, 
in analogy to the $\phi$ production 
via K$\bar{\rm K}$ \cite{Soff:1999et}.  

\section{Summary}
Here, we focused on the role of strong color fields that lead to 
(i) enhanced heavy flavor and diquark production probabilities, 
(ii) modified, shorter formation times,
(iii) stronger intrinsic transverse momenta. 
The effects on the $\phi$ meson include 
(a) enhanced yields, (b) a change of the composition of the spectra, 
(c) a hardening of the spectra, (d) modified baryon dynamics 
(hyperonization).
Moreover, we briefly discussed the competition of late stage vs.\ 
early production and the rescattering of the decay daughters. 
The differences between the kaonic and muonic $\phi$-mesons are desired! 

The quark molecular dynamics model provides a (simple) 
dynamical description of the hadronization process in the soft regime 
and needs to be further tested. 

Parton cascades are now technically established and provide 
a valuable tool to learn about the validity range of pQCD or 
necessary modifications. 

The study of excitation functions was shown to be essential to detect 
deficiencies in the models. 
Additional new developments include the investigation of 
multi-particle collisions, of charm in microscopic models, 
and the formulation of the initial conditions in a field 
theoretical language (parton ladders).

\vspace*{-0.4cm}
\ack
\vspace*{-0.3cm}
This work was supported by the Alexander von Humboldt-Foundation through 
a Feodor Lynen Fellowship, BMBF, DFG, GSI, and NSF grant PHY-03-11859.  
We thank S.\ Kesavan, V.\ Koch, J.~Randrup, S.~Scherer, 
H.\ St\"ocker, and N.\ Xu,  
for many helpful comments and/or for providing figures.  

\end{document}